# The Password Change Phase is Still Insecure

**Manoj Kumar**

Department of Applied Sciences and Humanities

Hindustan College of Science & Technology

Farah, Mathura, (U.P.) India-281122.

E. Mail: yamu_balyan@yahoo.co.in

**Abstract—** *In 2004, W. C. Ku and S. M. Chen proposed an efficient remote user authentication scheme using smart cards to solve the security problems of Chien et al.'s scheme. Recently, Hsu and Yoon et al. pointed out the security weaknesses of the Ku and Chen's scheme Furthermore, Yoon et al. also proposed a new efficient remote user authentication scheme using smart cards. Yoon et al. also modified the password change phase of Ku and Chen's scheme. This paper analyzes that password change phase of Yoon et al's modified scheme is still insecure.[1]*

## I. INTRODUCTION

A password based remote user authentication scheme is used to check the validity of a login request made by a remote user $U$ to gain the access rights on an *authentication server* (*AS*). In these schemes, the *AS* and the remote user $U$ share a secret, which is often called as password. With the knowledge of this password, the remote user $U$ uses it to create a valid login request to the *AS*. *AS* checks the validity of the login request to provide the access rights to the user $U$. Password authentication schemes with smart cards have a long history in the remote user authentication environment. So far different types of password authentication schemes with smarts cards [3] - [4] - [5] - [6] - [12] - [13] - [14] - [18] - [20] - [21] - [24] - [29] have been proposed.

[1] Manoj Kumar is with the Department of Applied Sciences and Humanities, Sharda Group of Institutions (*SGI*), Jawahar Nagar, Khandri, Agra, India - 282004, (e-mail: yamu_balyan@yahoo.co.in)

Lamport [17] proposed the first well-known remote password authentication scheme using smart cards. In Lamport's scheme, the *AS* stores a password table at the server to check the validity of the login request made by the user. However, high hash overhead and the necessity for password resetting decrease the suitability and practical ability of Lamport's scheme. In addition, the Lamport scheme is vulnerable to a small *n* attack [7]. Since then, many similar schemes [23]-[26] have been proposed. They all have a common feature: *a verification password table should be securely stored in the AS.* Actually, this property is a disadvantage for the security point of view. If the password table is stolen /removed /modified by the adversary, the *AS* will be partially or totally braked/affected.

In 2002, Chien–Jan–Tseng [13] introduced an efficient remote user authentication scheme using smart cards. In 2004, Ku and Chen [31] pointed out some attacks [7]-[28]-[30] on Chien – Jan and Tseng's scheme. According to Ku and Chen, Chien et al.'s scheme is vulnerable to a reflection attack [7] and an insider attack [30]. Ku and Chen claimed that Chien et al.'s scheme is also not reparable [28]. In addition, they also proposed an improved scheme to prevent the attacks: reflection attack and an insider attack on Chien–Jan–Tseng's scheme. In the same year, Hsu [10] pointed out that the Chien–Jan–Tseng's scheme is still vulnerable to a parallel session attack and Yoon et al. [11] claimed that the password change phase of improved scheme of Chien–Jan–Tseng's scheme is still insecure.

*A. Contributions*

This paper analyzes that the password change phase of Yoon et al.'s scheme is still insecure.

*B. Organization*

Section II reviews the Ku and Chen's scheme [31]. Section III reviews Hsu [10] and Yoon et al.'s comments on Ku and Chen's scheme .Section IV reviews Yoon et al.'s scheme [11]. Section V is about our observations on the security pitfalls in the password change phase of Yoon et al.'s scheme. Finally, comes to a conclusion in the section VI.

## II. REVIEW OF KU AND CHEN'S SCHEME

This section briefly describes Ku and Chen's scheme [31]. This scheme has four phases: the registration phase, login phase, verification phase and the password change phase. All these four phases are described below.

### A. Registration Phase

This phase is invoked whenever $U$ initially or re-registers to $AS$. Let $n$ denotes the number of times $U$ re-registers to AS. The following steps are involved in this phase.

Step.R1: User $U$ selects a random number $b$ and computes $PW_S = f(b \oplus PW)$ and submits her/his identity $ID$ and $PW_S$ to the $AS$ through a secure channel.

Step.R2: $AS$ computes a secret number $R = f(EID \oplus x) \oplus PW_S$, where $EID = (ID \| n)$ and creates an entry for the user $U$ in his account database and stores $n = 0$ for initial registration, otherwise set $n = n+1$, and $n$ denotes the present registration.

Step.R3: $AS$ provides a smart card to the user $U$ through a secure channel. The smart card contains the secret number $R$ and a one-way function $f$.

Step.R4: User $U$ enters his random number $b$ into his smart card.

### B. Login Phase

For login, the user $U$ inserts her/his smart card to the smart card reader and then keys the identity and the password to gain the access services. The smart card will perform the following operations:

Step.L1: Computes $C_1 = R \oplus f(b \oplus PW)$ and $C_2 = f(C_1 \oplus T_U)$. Here $T_U$ denotes the current date and time of the smart card reader.

Step.L2: Sends a login request $C = (ID, C_2, T_U)$ to the $AS$.

### C. Verification Phase

Assume $AS$ receives the message $C$ at time $T_S$, where $T_S$ is the current date and time at $AS$. Then the $AS$ takes the following actions:

Step.V1: If the identity $ID$ and the time $T_U$ are not valid, then $AS$ will rejects this login request.

Step.V2: Checks, if $C_2 \stackrel{?}{=} f(f(EID \oplus x) \oplus T_U)$, then the $AS$ accepts the login request and computes $C_3 = f(f(EID \oplus x) \oplus T_S)$. Otherwise, the login request $C$ will be rejected.

Step.V3: $AS$ sends the pair $T_S$ and $C_3$ to the user $U$ for mutual authentication.

Step.V4: If the time $T_S$ is invalid $i.e.$ $T_U = T_S$, then $U$ terminates the session. Otherwise, the user $U$ verifies the equation $C_3 \stackrel{?}{=} f(C_1 \oplus T_S)$ to authenticate $AS$.

### D. Password Change Phase

This phase is invoked whenever $U$ wants to change his password $PW$ with a new password, say $PW_{new}$. This phase has the following steps.

Step.P1: *U* inserts her/his smart card to the smart card reader keys the identity and the password and then requests to change the password. Next, *U* enters a new password $PW_{new}$.

Step.P2: *U*'s smart cards computes a new secret number $R_{new} = R \oplus f(b \oplus PW) \oplus f(b \oplus PW_{new})$ and then replaces *R* with $R_{new}$.

## III. REVIEW OF HSU AND YOON ET AL.'S COMMENT ON THE KU AND CHEN'S SCHEME

### A. Hsu's Comment

According to Hsu, Ku and Chen's scheme is vulnerable to a parallel session attack [10]. The intruder Bob intercepts the communication between the *AS* and user *U* and then from this intercepted information, he makes a valid login request to masquerade as a legal user. The intruder Bob applies the following steps for a successful parallel session attack.

✓ Intercepts the login request $C = (ID, C_2, T_U)$ which is sent by a valid user *U* to *AS*.

✓ Intercepts the response message $(C_3, T_S)$, which is sent by *AS* to he user *U*.

✓ Starts a new session with the *AS* by sending a fabricated login request $C_f = (ID, C_3, T_S)$.

The fabricated login request passes all the requirements for a successful authentication of the intruder Bob by the *AS*, due to the fact that the second part , $C_3$, of the login request also satisfies the verification equation $C_3 \overset{?}{=} f(f(EID \oplus x) \oplus T_S)$.

✓Finally, *AS* computes $C_4 = f(f(EID \oplus x) \oplus T_S)$ and responses with the message pair $(T^*_S, C_4)$ to the user *U* for mutual authentication, where is the current timestamp of the *AS*. Thus, the intruder intercepts and drops this message

### B. Yoon et al.'s Comment on Ku and Chen's Scheme

According to Yoon et al., the password change phase of Ku and Chen's scheme is insecure. When the smart card was stolen, an authorized user can easily replace the old password by a new password of her/his choice. First, the authorized user inters the smart card into the smart card reader, enters the identity *ID* and any password $PW^*$ of her/his choice and then requests to change the password. Next, the authorized user enters a new password $PW^*_{new}$ and then the smart card computes a new $R^*_{new} = R \oplus f(b \oplus PW^*) \oplus f(b \oplus PW^*_{new})$ and then replaces *R* with $R^*_{new}$, without any checking.

Thus, if the malicious user stole the user *U*'s smart card once, only for a small time and then change the valid password with an arbitrary password $PW^*$ , then the registered/ legal user *U* also will not be able to make a valid login request. The *AS* will not authenticate a registered user *U,* because $C_2 \neq f(f(EID \oplus x) \oplus T_U)$ in the verification phase.

## IV. YOON ET AL.'S SCHEME

This section briefly describes Yoon et al.'s scheme [11]. This scheme also has four phases: the registration phase, login phase, verification phase and the password change phase. All these four phases are described below.

### A. Registration Phase

This phase is invoked whenever *U* initially or re-registers to *AS*. Let *n* denotes the number of times *U* re-registers to AS. The following steps are involved in this phase.

❖ User *U* selects a random number *b* and computes $PW_S = f(b \oplus PW)$ and submits her/his identity *ID* and $PW_S$ to the *AS* through a secure channel.

❖ *AS* computes two secret numbers $V = f(EID \oplus x)$ and $R = f(EID \oplus x) \oplus PW_S$, where *EID* = $(ID \| n)$ and creates an entry for the user *U* in his account database and stores *n* = 0 for initial registration, otherwise set *n*= *n*+1, and *n* denotes the present registration.

❖ *AS* provides a smart card to the user *U* through a secure channel. The smart card contains two secret numbers *V*, *R* and a one-way function *f*.

❖ User *U* enters her/his random number *b* into his smart card.

### B. Login Phase

For login, the user *U* inserts her/his smart card to the smart card reader and then keys the identity and the password to gain access services. The smart card will perform the following operations:

❖ Computes $C_1= R \oplus f(b \oplus PW)$ and $C_2 = f(C_1 \oplus T_U)$. Here $T_U$ denotes the current date and time of the smart card reader.

❖ Sends a login request $C = (ID, C_2, T_U)$ to the *AS*.

### C. Verification Phase

Assume *AS* receives the message *C* at time $T_S$, where $T_S$ is the current date and time at *AS*. Then the *AS* takes the following actions:

❖ If the identity *ID* and the time $T_U$ is invalid *i.e.* $T_U = T_S$, then *AS* will rejects this login request.

❖ Checks, if $C_2 \overset{?}{=} f(f(EID \oplus x) \oplus T_U)$, then the $AS$ accepts the login request and computes $C_3$ = $f(f(EID \oplus x) \oplus T_S)$. Otherwise, the login request $C$ will be rejected.

❖ $AS$ sends the pair $T_S$ and $C_3$ to the user $U$ for mutual authentication.

❖ If the time $T_S$ is invalid *i.e.* $T_U = T_S$, then $U$ terminates the session. Otherwise, $U$ verifies the equation $C_3 \overset{?}{=} f(C_1 \oplus T_S)$ to authenticates $AS$.

### D. Password Change Phase

This phase is invoked whenever $U$ wants to change his password $PW$ with a new one, say $PW_{new}$. This phase has the following steps.

❖ $U$ inserts her/his smart card to the smart card reader and then keys her/his identity and the old password $PW$ and then requests to change the password.

❖ $U$'s smart cards computes $V^* = R \oplus f(b \oplus PW)$.

❖ Compare this calculated value $V^*$ with the secret value $V$, which is stored in the smart card of the user $U$. If they are equal, then $U$ can select a new password $PW_{new}$, otherwise the smart card rejects the password change request.

❖ $U$'s smart cards computes a new secret number $R_{new} = V^* \oplus f(b \oplus PW_{new})$ and then replaces $R$ with $R_{new}$.

## V. SECURITY ANALYSIS OF THE PASSWORD CHANGE PHASE OF YOON ET AL.'S SCHEME

Although, the password change phase of Ku and Chen's scheme is modified by Yoon et al. [11] to remove its security weaknesses. But, we analyze that the modified password phase of Yoon et al.' scheme is still not secure. This section discusses the security weaknesses of the password change phase of Yoon et al.'s scheme and proves that the modified phase is still vulnerable to security attack.

### A. Security weaknesses in the Password Change Phase against the Outsiders

Observe the password change phase of Yoon el al.'s scheme, to replace/change the old password $PW$ with a new password $PW_{new}$, the user/performer should be in possession of the old password $PW$. The following section describes how any outsider /malicious user can recover the password $PW$ first and then apply this peace of information to make for the success of her/his attack.

It is clear that the smart card of a legal user $U$ in Yoon et al.'s scheme contains: *the secret value V, R,* and *a random number b and a public hash function f.* According to Kocher et al. [22] and Messerges et al. [29], for the security point of view, to store the secret information in smart cards is not a good practice. On the basis of these assumptions [22]-[29], an antagonist is able to breach the secrets $V$, $R$ and $b$, which are stored in the smart card of the user and then he will be able to perform a password guessing attack to obtain the password. For the success of this attack, by using the breached secrets $R$ and $b$ the adversary will perform the following operations:

- *The antagonist intercepts the login request $C = (ID, C_2, T_U)$ and guesses a password $PW^*$.*

- *Computes $C_1^* = R \oplus f(b \oplus PW^*) = f(ID \oplus x)^*$ and $C_2^* = f(C_1^* \oplus T_U)$.*

- *Checks if $C_2^* \stackrel{?}{=} C_2$, then the adversary has correctly guessed the password $PW^* = PW$ and $C_1^* = C_1$. Otherwise, the adversary goes to* step: 1.

Once the adversary has correctly obtained $C_1$, instantly, the password $PW^*$ corresponding to $C_1$ will be the correct password and then successfully, he can change the password of the user $U$. Consequently, when the smart card was stolen, the antagonist is able to recover the password $PW$ of the user and once the adversary has correctly obtain the password $PW$, then he will be able to destruct anything of his choice. Since our focus and aim is to show that the password change phase of Yoon et al.'s scheme, which is shown below that an authorized user ( antagonist) can easily replace the old password $PW$ by a new password of her/his choice. For the success, the antagonist applies the following actions.

- Inters the smart card into the smart card reader, enters the identity $ID$ and any password $PW$ and then requests to change the password.

- The smart card of the user computes $V^* = R \oplus f(b \oplus PW)$ and then compare the computed value $V^*$ with the stored value $V$. Obviously, both the value will be the same, because the adversary has entered the correct password. In this way, the smart card accepts the password change request.

- Selects a new password $PW^*_{new}$ and supplies it to the smart card reader and ultimately the smart card computes a new $R^*_{new} = R \oplus f(b \oplus PW) \oplus f(b \oplus PW^*_{new})$ and then replaces $R$ with $R^*_{new}$.

Thus, if the malicious user stole the user $U$'s smart card she/he will be able to make a destructive action of her/his choice. Thus, the adversary is able to change the password with a new password of his/his choice. Now the registered/ legal user $U$ also will not be able to make a valid login request with her/his valid smart card because now the her/his old password $PW$ will not work .

### B. Security weaknesses in the Password Change Phase against the Insider

This section proves that the password change phase of Yoon et al.'s scheme is not secure against an antagonist insider at AS. *In Yoon et al.'s scheme, observe the registration phase, the User U selects a random number b and computes $PW_S = f(b \oplus PW)$ and submits her/his identity ID and $PW_S$ to the AS through a secure channel. It means the insider of AS is in possession of the number $PW_S = f(b \oplus PW)$ for the legal user U. Again the AS computes two secret numbers $V = f(EID \oplus x)$ and $R = f(EID \oplus x) \oplus PW_S$, where $EID = (ID \| n)$. Thus, the insider of AS is also in possession of the secret numbers V and R for the legal user U.*

Suppose the user *U* is using the same password *PW* continuously, which is supplied by the *AS* at the time of registration, then the insider at *AS* will be able to change the password *PW* with a new password of her/his choice. If the smart card is in possession of an antagonist insider at *AS* for short time, then first, the insider inters the smart card into the smart card reader and can directly supply the value *V* to the smart card reader. Either, he directly supplies *V* or in place of $f(b \oplus PW)$, he supplies the value $PW_S$ without using the hash button. Next, the antagonist insider enters a new password $PW^*_{new}$ and then the smart card computes a new $R_{new} = R \oplus f(b \oplus PW) \oplus f(b \oplus PW^*_{new})$ and then replaces *R* with $R_{new}$.

Thus, if the malicious insider stole the user *U*'s smart card once, only for a small time and then he can replace the user's password forever in such a way that the user *U* also will not be able to make a valid login request with her/his valid smart card because now the her/his old password *PW* will not work properly.

Thus the Yoon et al.'s password change phase is still insecure and that is under the threat of poor reparability.

## VI. Conclusion

This paper analyzed that security weaknesses still exist in the password change phase of modified scheme of Yoon et al.'s scheme. The password change phase is still vulnerable to security attacks by an outsider as well as an antagonist insider at *AS*. Thus, the security pitfalls still exist in Yoon et al.'s scheme.

### References

[1]   A. J. Menezes, P. C. vanOorschot and S. A. Vanstone, *Handbook of Applied Cryptography*, pp. 490 – 524, 1997.

[2] C. C. Chang and K. F. Hwang, "Some forgery attack on a remote user authentication scheme using smart cards," *Informatics,* vol. 14, no. 3, pp. 189 - 294, 2003.

[3] C. C. Chang and S. J. Hwang, "Using smart cards to authenticate remote passwords," *Computers and Mathematics with applications,* vol. 26, no. 7, pp. 19-27, 1993.

[4] C. C. Chang and T. C. Wu, "Remote password authentication with smart cards," *IEE Proceedings-E,* vol. 138, no. 3, pp. 165-168, 1993.

[5] C. C. Lee, L. H. Li and M. S. Hwang, "A remote user authentication scheme using hash functions," *ACM Operating Systems Review,* vol. 36, no. 4, pp. 23-29, 2002.

[6] C. C. Lee, M. S. Hwang and W. P. Yang, "A flexible remote user authentication scheme using smart cards," *ACM Operating Systems Review,* vol. 36, no. 3, pp. 46-52, 2002.

[7] C. J. Mitchell and l. Chen, "Comments on the S/KEY user authentication scheme," *ACM Operating System Review,* vol. 30, No. 4, pp. 12-16, Oct 1996.

[8] C. K. Chan and L. M. Cheng, "Cryptanalysis of a remote user authentication scheme using smart cards," *IEEE Trans. Consumer Electronic,* vol. 46, no. 4, pp. 992-993, 2000.

[9] C. Mitchell, "Limitation of a challenge- response entity authentication," Electronic Letters, vol. 25, No.17, pp. 1195- 1196, Aug 1989.

[10] C.L Hsu, "Security of Chien et al.'s remote user authentication scheme using smart cards," *Computer Standards and Interfaces,* vol. 26, no. 3, pp. 167 - 169, 2004.

[11] E. J. Yoon, E. K. Ryu and K. Y. Yoo, Further improvement of an efficient password based remote user authentication scheme using smart cards", *IEEE Trans. Consumer Electronic,* vol. 50, no. 2, pp. 612-614, May 2004.

[12] H. M. Sun, "An efficient remote user authentication scheme using smart cards," *IEEE Trans. Consumer Electronic,* vol. 46, no. 4, pp. 958-961, Nov 2000.

[13] H. Y. Chien, J.K. Jan and Y. M. Tseng, "An efficient and practical solution to remote authentication: smart card," *Computer & Security,* vol. 21, no. 4, pp. 372-375, 2002.

[14] J. J. Shen, C. W. Lin and M. S. Hwang, "A modified remote user authentication scheme using smart cards," *IEEE Trans. Consumer Electronic,* vol. 49, no. 2, pp. 414-416, May 2003.

[15] K. C. Leung, L. M. Cheng, A. S. Fong and C. K. Chen, "Cryptanalysis of a remote user authentication scheme using smart cards", *IEEE Trans. Consumer Electronic,* vol. 49, no. 3, pp. 1243-1245, Nov 2003.

[16] L. H. Li, I. C. Lin and M. S. Hwang, "A remote password authentication scheme for multi-server architecture using neural networks," *IEEE Trans. Neural Networks,* vol. 12, no. 6, pp. 1498-1504, 2001.

[17] L. Lamport, "Password authentication with insecure communication," *communication of the ACM,* vol. 24, no. 11, pp. 770-772, 1981.

[18] M. Kumar, "New remote user authentication scheme using smart cards," *IEEE Trans. Consumer Electronic,* vol. 50, no. 2, pp. 597-600, May 2004.

[19] M. Kumar, "Some remarks on a remote user authentication scheme using smart cards with forward secrecy." *IEEE Trans. Consumer Electronic,* vol. 50, no. 2, pp. 615-618, May 2004.

[20] M. S. Hwang and L. H. Li, "A new remote user authentication scheme using smart cards," *IEEE Trans. Consumer Electronic,* vol. 46, no. 1, pp. 28-30, Feb 2000.

[21] M. Udi, "A simple scheme to make passwords based on the one-way function much harder to crack," *Computer and Security,* vol. 15, no. 2, pp. 171 - 176, 1996.

[22] P. Kocher, J. Jaffe and B. Jun, "Differential power analysis," *Proc. Advances in Cryptography* (CRYPTO'99), pp. 388-397, 1999.

[23] R. E. Lennon, S. M. Matyas and C. H. Mayer, "Cryptographic authentication of time-variant quantities." *IEEE Trans. on Commun.,COM* -29, no. 6 , pp. 773 - 777, 1981.

[24] S. J. Wang, "Yet another login authentication using N-dimensional construction based on circle property," *IEEE Trans. Consumer Electronic,* vol. 49, No. 2, pp. 337-341, May 2003.

[25] S. M. Yen and K.H. Liao, "Shared authentication token secure against replay and weak key attack," *Information Processing Letters,* pp. 78-80, 1997.

[26] T. C. Wu, "Remote login authentication scheme based on a geometric approach," *Computer Communication,* vol. 18, no. 12, pp. 959 - 963, 1995.

[27] T. ElGamal, "A public key cryptosystem and a signature scheme based on discrete logarithms," *IEEE Trans. on Information Theory,* vol. 31, No. 4, pp. 469-472, July 1985.

[28] T. Hwang and W.C. Ku, "Reparable key distribution protocols for internet environments," *IEEE Trans. Commun. ,* vol. 43, No. 5, pp. 1947-1950, May 1995.

[29] T. S. Messerges, E. A. Dabbish and R. H. Sloan, " Examining smart card security under the threat of power analysis attacks," *IEEE Trans. on Computers,* vol. 51, no. 5, pp. 541 –552, May 2002.

[30] W. C. Ku, C. M. Chen and H. L. Lee, " Cryptanalysis of a variant of Peyravian- Zunic's password authentication scheme," *IEICE Trans. Commun,* vol. E86- B, no. 5, pp. 1682 –1684, May 2002.

[31] W. C. Ku and S. M. Chen, " Weaknesses and improvements of an efficient password based user authentication scheme using smart cards," *IEEE Trans. Consumer Electronic,* vol. 50, no. 1, pp. 204 –207, Feb 2004.

[32] Y. L. Tang, M. S. Hwang and C. C. Lee, "A simple remote user authentication scheme," *Mathematical and Computer Modeling,* vol. 36, pp. 103 - 107, 2002.

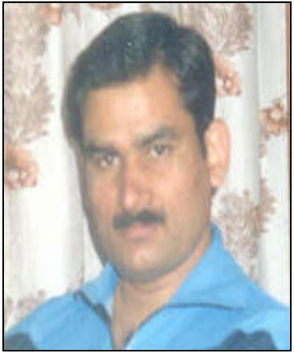

**Manoj Kumar** received the B.Sc. degree in mathematics from Meerut University Meerut, in 1993; the M. Sc. in Mathematics (Goldmedalist) from C.C.S.University Meerut, in 1995; the M.Phil. (Goldmedalist) in *Cryptography*, from Dr. B.R. A. University Agra, in 1996; submitted the Ph.D. thesis in *Cryptography*, in 2003. He also taught applied Mathematics at D.A.V. College, Muzaffarnagar, India from Sep, 1999 to March, 2001; at S.D. College of Engineering & Technology, Muzaffarnagar, U.P., India from March, 2001 to Nov, 2001; at Hindustan College of Science & Technology, Farah, Mathura, continue since Nov, 2001. He also qualified the *National Eligibility Test* (NET), conducted by *Council of Scientific and Industrial Research* (CSIR), New Delhi- India, in 2000. He is a member of Indian Mathematical Society, Indian Society of Mathematics and Mathematical Science, Ramanujan Mathematical society, and Cryptography Research Society of India. His current research interests include Cryptography, Numerical analysis, Pure and Applied Mathematics.